\documentclass[prl,twocolumn,showpacs,aps,superscriptaddress]{revtex4}
\usepackage{graphicx,epsfig}
\usepackage{rotate}
\usepackage{color}
\usepackage{amsmath,amssymb}

\newcommand{\dis}{\displaystyle}

\newcommand{\mb}{\mathbf}
\newcommand{\mr}{\mathrm}

\newcommand{\ba}{\begin{array}}
\newcommand{\ea}{\end{array}}

\begin{document}

\title{Communicability reveals a transition to coordinated behavior in multiplex networks}

\author{E. Estrada}
\affiliation{Department of Mathematics \& Statistics, Institute of Complex Systems, University of Strathclyde, Glasgow G1 1HX, UK}

\affiliation{The Institute of Quantitative Theory and Methods, Emory University, GA 30033, USA, }

\author{J. G\'omez-Garde\~nes}
\affiliation{Institute for Biocomputation and Physics of Complex Systems (BIFI), Universidad de Zaragoza, 50018 Zaragoza, Spain}
\affiliation{Departamento \ de F\'{\i}sica de la Materia Condensada, Universidad de Zaragoza, 50009 Zaragoza, Spain}

\begin{abstract}
We analyse the flow of information in multiplex networks by means of the communicability function. First, we generalize this measure from its definition from simple graphs to multiplex networks. Then, we study its relevance for the analysis of real-world systems by studying a social multiplex where information flows using formal/informal channels and an air transportation system where the layers represent different air companies. Accordingly, the communicability, which is essential for the good performance of these complex systems, emerges at a systemic operation point in the multiplex where the performance of the layers operates in a coordinated way very differently from the state represented by a collection of unconnected networks.
\end{abstract}


\maketitle

Complex systems have been usually considered as ensembles of entities whose interactions are encoded in the form of a complex network \cite{albert,newman,boccaletti,dorogovtsev}. This approach neglects the fact that in real-world complex systems agents usually interact simultaneously in many diverse ways. Very recently, the study of multiplex networks has attracted a great deal of attention in the literature \cite{kurant,cardillo,szell,mucha}. In a multiplex, every entity of the complex system is split into $h$  layers, each representing a different kind of interaction among these agents. This kind of complex system representation is very convenient for the analysis, among others, of socio-economic and of transportation systems, where the layers represent different social communication or transportation channels.

The recent interest in multiplex networks has been focused on the characterization of their structural properties \cite{cardillo,mucha,sole-ribalta,dedomenico,nicosia,centrality,clustering} and the associated critical phenomena \cite{gomez-gardenes,baxter,cozzo,bianconi,gomez,granell,banos}. The latter ones arise as a consequence of having different dynamical processes taking place simultaneously within each of the networked layers of the multiplex. Perhaps the most interesting aspect of this research is to unveil how the combination of different physical properties of each network layer yields new emergent behaviors that cannot be understood as the simple sum of the properties of each networked component. 
For instance, in \cite{dedomenico,radicchi} it has been found out that multiplexes display a transition from a regime in which the system behaves as a set of independent networks to the one in which a coordinated behavior emerges. These transitions are obtained by decreasing the relative importance of the connections between the agents in each of the layers in relation to those representing the flow between the layers.

In this letter, we analyze how the communication among the nodes in certain multiplex complex systems is affected by the coupling between the different layers. This analysis is carried by means of a generalization of the communicability function \cite{estrada,estrada2,estrada3} to multiplex networks. The communicability function quantifies the number of possible routes that two nodes have to communicate with each other.
We then show that communicability unveils the transition from a small coupling regime, when the multiplex behaves just as a collection of individual networks, to the one in which it acts in a coordinated way.

{\it Communicability in multiplexes.} Let us consider a multiplex formed by $h$  layers designated $L_1,\ldots,L_h$ (see Fig.~1), and their respective adjacency matrices by  $\mb{A}_1,\ldots,\mb{A}_h$. The multiplex matrix is then given by  $\mb{M} = \mb{A}_L + \mb{C}_{LL}$, where $\mb{A}_L$  is the adjacency matrices of the two layers, $\mb{A}_L = \dis\oplus^h_{i=1} \mb{A}_i$, and  $\mb{C}_{LL}$ is a matrix describing the interlayer interaction
\setcounter{equation}{2}
\begin{equation}
\mb{C}_{LL} = \left(\ba{llcl}\mb{0}&\mb{C}_{12}&\cdots&\mb{C}_{1h}\\ \mb{C}_{21}&\mb{0}&\cdots&\mb{C}_{2h}\\ \vdots&\vdots&\ddots&\vdots\\ \mb{C}_{h1}&\mb{C}_{h2}&\cdots&\mb{0}\ea\right),
\end{equation}
where $\mb{C}_{ij}$  represents the interaction of layer $i$  with layer $j$. Here we consider $\mb{C}_{ij} = \mb{C}_{ji} = \mb{C} = \omega I$, for all layers $i$  and  $j$,  $\omega$ is a parameter describing the strength of the interlayer interaction and  $\mb{I}$ is the corresponding identity matrix. In this case  $\mb{C}_{LL} = \mb{C} \otimes (\mb{E} - \mb{I})$, where $\mb{E}$  is an all-ones  $h\times h$ matrix.

\begin{figure}[t]
\begin{center}
\includegraphics[width=0.275\textwidth]{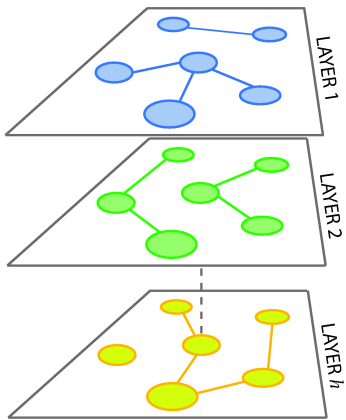}
\end{center}
\caption{Illustration of a multiplex formed by $h$ network layers. Each layer is composed of $N=6$ nodes and each of the nodes is represented in each of the layers. The connectivity of the nodes is, in principle, different in each layer of the multiplex. Apart from the connections that a node shares within each layer we consider that a node is also connected with each of its representations in the remaining network layers.}
\label{fig:multiplex}
\end{figure}

Here we are interested in accounting for all the walks between any pair of nodes in the multiplex. It is known that the number of walks of length $k$  between the nodes  $p$ and $q$  in a network is given by the  $p,q$-entry of the  $k$th power of the adjacency matrix of the network. Consequently, the walks of $k$  length in the multiplex are given by the different entries of  $\mb{M}^k$. In principle, the walks can contain hops of two different kinds, {\em i.e.}, intra-layer and inter-layer hops. 
Following the definition of the communicability in simple networks, we are interested in giving more weight to the shortest walks than to the longer one. Consequently, we define the communicability between two nodes  $p$  and  $q$ in the multiplex as a weighted sum of all walks from $p$  to $q$ in the following way:
\begin{equation}
G_{pq} = \mb{I} + \mb{M} + \frac{\mb{M}^2}{2!} + \cdots = \sum^\infty_{k=0} \, \frac{\mb{M}^k}{k!}\,.
\end{equation}
Consequently, the communicability between the nodes $p$  and  $q$ in the multiplex is given by
\begin{equation}
G_{pq} = [\exp (\mb{A}_L + \mb{C}_{LL})]_{pq}.
\end{equation}
Obviously, if  $\omega = 0$ all the interlayer communication is knocked out and
\begin{equation}
G_{pq} = \left(\ba{ccc}\exp(\mb{A}_1)&&0\\
&\ddots&\\
0&&\exp(\mb{A}_h)\ea\right)_{pq},
\end{equation}
so that communicability is exactly equal to that of a collection of independent networks.
In order to quantify the total amount of communicability broadcasted and received by a given node in the multiplex we consider the following approach. Let
\begin{equation}
\mb{G} = \exp(\mb{A}_L + \mb{C}_{LL}) = \left(\ba{llcl}\mb{G}_1&\mb{G}_{12}&\cdots&\mb{G}_{1h}\\ \mb{G}_{21}&\mb{G}_2&\cdots&\mb{G}_{2h}\\\vdots&\vdots&\ddots&\vdots\\ \mb{G}_{h1}&\mb{G}_{h2}&\cdots&\mb{G}_1\ea\right),
\end{equation}
where $\mb{G}_i$  is the matrix representing the communicability between every pair of nodes in the layer $i$  of the multiplex. It is important to note that $\mb{G}_i \ne \exp (\mb{A}_i)$  due to the coupling between the layers. Then, the communicability broadcasted, respectively received, by the node $p$  in the $k$th layer of the multiplex is
 \begin{eqnarray}
 G_k^{\mr{broadcast}} (p) &=& \sum^N_{q=1} \mb{G}_k (q, p),\\
 G_k^{\mr{received}} (p) &=& \sum^N_{q=1} \mb{G}_k (p, q).
 \end{eqnarray}
Notice that these indices contain information about both the intra- and inter-layer walks. If all layers are symmetric, {\em i.e.}, undirected networks,  $G_k(p) = G_k^{\mr{broadcast}} (p) = G_k^{\mr{received}}(p)$.

In order to account for the mean broadcasting and receiving activity of a node in the multiplex we consider the following approaches. First we assume that the information between the nodes $p$  and $q$  is flowing in parallel at the different layers. Then, we consider that the mean information broadcasted or received by a node $p$  is accounted for by the harmonic mean  $H^{\rm{type}}(p)$ (type = broadcast, receive) of the communicability (broadcasted or received) by this node in all the layers of the multiplex. 
In addition, to compare the results we use the aggregate network $\hat G$  defined as follows. Let $G_1 = (V_1, E_1), G_2 = (V_1, E_2), \ldots, G_k = (V_1, E_k)$  be the set of layers of the multiplex. Then, $\hat G = (\hat V, \hat E)$   where $\hat V = V_1$  and $\hat E = \dis\cup^k_{i=1} E_i$.

{\it Communicability in a social multiplex.} In most socio-economical organizations there is a formal or official structure, which defines the official hierarchy, lines of authority and of communication. In parallel, there is a network of friendships that tie people together in ways that have nothing to do with the official structure. This situation is very clear in a social multiplex obtained as the result of 16 months of observation of an office politics \cite{thurman}. The office is formed by 15 members of an overseas branch of a large international organization. This multiplex is formed by two layers, the first layer corresponds to a directed network comprising the formal organizational chart of the employees, whereas the second layer represents the informal association among the employees. The multiplex network is represented in Fig.~2.

\begin{figure}[t]
\begin{center}
\includegraphics[width=0.47\textwidth]{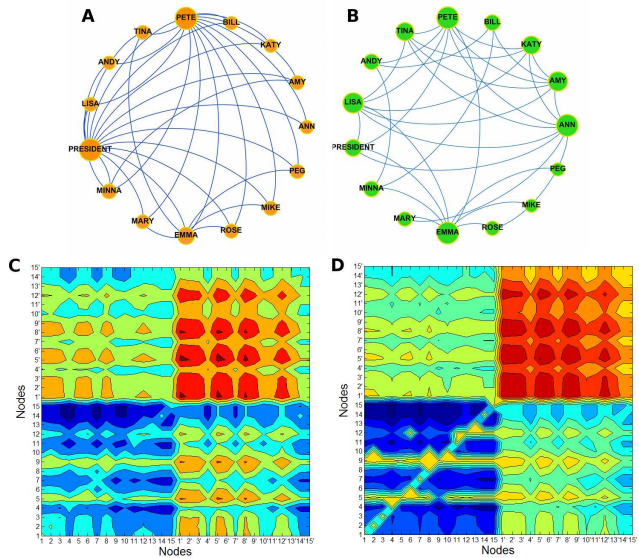}
\end{center}
\caption{In the top we show the formal (left) and informal (right) communication layers among the 15 members of the organization studied in \cite{thurman}. The formal layer of communication forms a directed network while the informal one, representing friendship ties, is undirected. In the bottom part we show two contour plots of the communicability between the 15 members for two different coupling constants between the formal and informal layers of communication: $\omega=0.1$ (left) and $\omega=1$ (right). The indexes of the matrices correspond to: 1: ANN, 2: AMY, 3: KATY, 4: BILL, 5: PETE, 6: TINA, 7: ANDY, 8: LISA, 9: PRESIDENT, 10: MINNA, 11: MARY, 12: EMMA, 13: ROSE, 14: MIKE, 15: PEG.}
\label{fig:multiplex}
\end{figure}

During this period two employees, Emma and Minna, were the targets of a leveling coalition formed by 6 members of staff. From a network perspective the identification of the attacking coalition is not difficult as their members form a clique in the informal social layer of the multiplex. This coalition is formed by Ann, Katy, Amy, Pete, Tina and Lisa. The analysis of the communicability in the informal layer of the multiplex also reveals the importance of this coalition in the diffusion of information in the network. In Table 1 it can be seen that the six members of the coalition are the highest broadcasters of information in this layer in agreement with the observation made by Thurman that [22]: ``{\it Within the network a large number of rumors circulated rapidly among Pete, Ann, Amy, Katy, Tina, and Lisa}.'' However, nothing is evident about the victims of the attack from the analysis of the separated layers. In the informal layer of communication, Emma occupies the position immediately after the attacking coalition in the ranking of broadcasted communicability. However, Minna only appears at the bottom three of the ranking together with Mike and Peg. In the formal layer there are only four broadcasters: Pete, the President, Emma and Minna. We recall that Emma had been promoted to administrative manager and Minna was also in a managerial position. However, neither the communicability at the formal nor at the informal layer reveals any hint about the plausible causes for the attacks. On the other hand, in the aggregate network the ranking of the employees according to their broadcasted communicability is mixed up and while Emma is the fifth in broadcasting information, Minna occupies the position number nine.

The communicability between every pair of employees in the office for the two layers (administrative and informal) is given in the bottom part of Fig.~2 for two different values of the strength of the interlayer interaction $\omega=0.1$ (left) and $\omega=1$ (right). It can be seen that most of the communication flow takes place on the informal layer of the multiplex.

\begin{table}
\begin{center}
\begin{tabular}{lccc}\hline &Formal&Informal&Aggregate\\
\hline
PETE&17.50&436.52&1137.76\\
ANN&1.00&414.39&873.45\\
AMY&1.00&353.47&682.67\\
KATY&1.00&337.97&652.99\\
TINA&1.00&337.97&652.99\\
LISA&1.00&419.20&909.02\\
EMMA&6.00&274.51&776.81\\ \hline
MINNA&3.00&88.97&442.81\\ \hline
PRESIDENT&26.17&268.40&1137.76\\
BILL&1.00&99.35&218.13\\
ANDY&1.00&111.75&279.47\\
MARY&1.00&121.19&254.66\\
ROSE&1.00&121.19&254.66\\
MIKE&1.00&49.94&120.99\\
PEG&1.00&49.94&120.99
\end{tabular}
\end{center}
\caption{Broadcasted communicability in the formal and informal layer of communication for the 15 members of the social multiplex. We show the case of zero coupling between the two layers ($\omega=0$) as well as the case of the aggregate network.}
\end{table}

We consider now the harmonic mean of the communicability broadcasted in both layers for different values of the coupling constant (see Table 2). When the coupling between the formal and informal layers is relatively weak ($0.1 \le \omega < 0.5$) Emma and Minna occupy a privileged position in their broadcasting communicability, which place them only after Pete and the President and well over the rest of the members of the attacking coalition, who at the same time are better broadcasters than the rest of the employees. As the coupling constant increases the informal communication layer receives more importance in determining the amount of information broadcasted. In this scenario, Minna starts to loss their hierarchy in broadcasting information and she passes from being the 4th broadcaster at $0.1 \le \omega < 0.5$  to the 9th place for  $\omega \ge 0.7$. This situation might explain why Emma and Minna have been the object of the coalition attacks. The coalition, which is very well communicated at the informal level, could see in Emma and Minna as a thread to their position as major broadcasters or controllers of the information flow in the office. This, of course, would never happen if the employees consider the informal level of communication only. But the coupled communication between the two layers, which the actors of the network would perceive as a unique block in which the information is propagated, well justify the feeling of this thread. This example clearly illustrates how neither the isolated layers nor the aggregate network can explain the ways in which information flows in a multiplex and affects its nodes. 

\begin{table}
\begin{center}
\begin{tabular}{lccccc} \hline
&$\omega = 0.1$&$\omega = 0.3$&$\omega = 0.5$&$\omega = 0.8$&$\omega = 1.0$\\ \hline
PETE&34.47&41.01&54.13&86.51&117.06\\
ANN&2.36&5.30&11.27&26.27&40.73\\
AMY&2.30&4.71&9.61&21.92&33.78\\
KATY&2.28&4.58&9.25&20.98&32.29\\
TINA&2.28&4.58&9.25&20.98&32.29\\
LISA&2.37&5.38&11.49&26.85&41.67\\ \hline
EMMA&12.12&15.16&21.28&36.56&51.15\\
MINNA&6.00&7.55&10.66&18.29&25.44\\ \hline
PRESIDENT&48.44&54.47&66.51&95.88&123.32\\
BILL&2.07&2.83&4.35&8.18&11.86\\
ANDY&2.10&3.02&4.88&9.58&14.10\\
MARY&2.10&3.01&4.87&9.53&14.01\\
ROSE&2.10&3.01&4.87&9.53&14.01\\
MIKE&2.02&2.47&3.39&5.69&7.91\\
PEG&2.02&2.47&3.39&5.69&7.91
\end{tabular}
\end{center}
\caption{Harmonic means of the received and broadcasted communicability for the 15 members of the social multiplex studied for different values of the coupling constant $\omega$.}
\end{table}

{\it Communicability in an airports multiplex.} Information, generally speaking, not only flows across the multiple layers of social systems. Airport transportation networks also represent an excellent example of a coupled multiplex system. Here we consider 450 European airports and 6 airlines, subdivided into {\it major} or {\it traditional} (British Airways, Lufthansa and AirFrance) and {\it low-cost} fares (Easyjet, AirBerlin and Ryanair). Each layer represents the air connectivity between the 450 airports provided by the corresponding airline \cite{cardillo}. The networks in each layer are undirected as if there is a flight from A to B, there is always a returning flight from B to A.

\begin{table*}
\begin{center}
\begin{tabular}{rllll} \hline
Rank&$\omega = 0.0$&$\omega = 0.1$&$\omega = 1.0$&Aggregate\\ \hline
1&Paris CdG&London Stansted&London Stansted&Frankfurt\\
2&Barcelona&Madrid&Dublin&Munich\\
3&Venice&Barcelona&Madrid&London Stansted\\
4&Amsterdam&Paris CdG&Palma de Mallorca&London Gatwick\\
5&Copenhagen&Dublin&Bergamo&Larnaca\\
6&Madrid&Malaga&Alicante&D\"usseldorf\\
7&Frankfurt&Bergamo&Barcelona&Madrid\\
8&Prague&Palma de Mallorca&Malaga&Paris CdG\\
9&Athens&Venice&Brussels South&Palma de Mallorca\\
10&Tolouse-Blagnac&Alicante&Pisa&Barcelona
\end{tabular}
\end{center}
\caption{Ranking of European airports on the basis of their harmonic mean of communicability for different coupling constants in the multiplex and for the aggregate network.}
\end{table*}

Our main goal here is to study how airport centrality, in terms of communicability, emerges from the coupling between the layers in the multiplex. We start by studying the harmonic $H(i)$  means of the communicability $G_k(i)$  of each airport in the respective layer for different values of the coupling constant. When there is no coupling between the layers, i.e.,  $\omega = 0.0$, $H(i)$  represents the harmonic mean of the communicability in each isolated layer or airline. As can be seen in Table 3 these airports are mainly the bases for major airline companies, such as Paris Charles de Gaulle (AirFrance) or those with the presence of most of the six airlines studied. In fact, if we consider the two harmonic means for the communicability in the major $H^{\rm{major}} (i)$  and low-cost $H^{\rm{low-cost}}(i)$  companies, respectively, we observe that the communicability in the uncoupled networks is dominated by major companies. For instance, the Pearson correlation coefficient between $H(i)$  and $H^{\rm{major}}(i)$  is 0.76, while that for $H(i)$  and $H^{\rm{low-cost}}(i)$  is only 0.29.

As soon as some coupling between the layers is allowed a different picture starts to emerge. For a small coupling constant, such as  $\omega = 0.1$, a few new airports show up as the most central ones in terms of their communicability. For instance, the London Stansted and Dublin airports now appear among the top ten most central airports in terms of their communicability. These airports are the main bases for low-cost fare companies such as Ryanair. Among the companies studied, Ryanair also has the largest presence in the airport of Madrid Barajas, which now occupies the second place in the ranking. When the coupling between the layers in the multiplex increases further, such as to  $\omega = 1.0$, these three latter airports become the most central ones. However, this increment in the relevance of these airports with heavy presence of low-cost companies is not developed in detriment of the role played by major airlines.

If we consider the correlation coefficient between $H(i)$  and $H^{\rm{major}}(i)$  for the coupling $\omega = 0.1$  it is 0.97 and that for  $H(i)$ vs.  $H^{\rm{low-cost}}(i)$ has also increased up to 0.65. For the coupling constant $\omega = 1.0$  these correlations have increased to 0.99 and 0.98, respectively. That is, the increase in the coupling between the different layers in the multiplex equilibrates the role played by major and low-cost companies in determining the centrality of the respective airports. In other words, by coupling with a moderate strength the airlines of the multiplex a situation in which major and low-cost airlines operate in a coordinated way shows up. However, the coupling for this balanced regime has to be moderate enough since increasing more $\omega$  we approach the aggregate network. In this case the correlation coefficients between $H(i)$  and $H^{\rm{major}}(i)$  and $H(i)$  and $H^{\rm{low-cost}}(i)$   have dropped to 0.7 and 0.87 pointing out a less equilibrated regime than that for moderate values for inter-layer coupling.

{\it Conclusion.} In this work we have analyzed the flow of information in multiplex networks by means of their communicability. After generalizing this measure from the case of simple networks to the most realistic scenario of multiplex networks, we have studied its relevance in two real systems. The first represents a small social multiplex formed by individuals in an organization, in which the information flows across a formal layer reflecting the hierarchical structure of the organization and another one representing the informal ways of communication among the actors. The second multiplex represents the European Air-transportation system in which air traffic between European airports is operated by 6 air-companies. Our study points out that the communicability, being essential for the good performance of these two real systems, shows the difference between a collection of unconnected networks and the systemic operation point in which the performance of the layers operates in coordinated way. In both cases the multiplex nature of the systems is essential to explain the flow of information and the centrality of nodes different to the simplistic limits in which the networked layers are disconnected or aggregated.

\acknowledgments This work has been partially supported by the Spanish MINECO under projects FIS2011-25167 and FIS2012-38266-C02-01; and by the European FET project MULTIPLEX (317532. J.G.G. is supported by the MINECO through the Ram\'on y Cajal Program.

\end{document}